\documentclass[conference]{IEEEtran}
\IEEEoverridecommandlockouts
\usepackage{diagbox}
\usepackage{cite}
\usepackage{amsmath,amssymb,amsfonts}
\usepackage{algorithmic}
\usepackage{graphicx}
\usepackage{subcaption}
\usepackage{caption}
\usepackage{textcomp}
\usepackage{array}
\usepackage{xcolor}
\usepackage{url}
\usepackage{tikz}

\newcommand\copyrighttext{%
  \footnotesize \copyright 2021 IEEE. Personal use of this material is permitted. 
  Permission from IEEE must be obtained for all other uses, in any current or future media, 
  including reprinting/republishing this material for advertising or promotional purposes,
  creating new collective works, for resale or redistribution to servers or lists, 
  or reuse of any copyrighted component of this work in other works.}
\newcommand\copyrightnotice{%
\begin{tikzpicture}[remember picture,overlay]
\node[anchor=south,yshift=10pt] at (current page.south) {\fbox{\parbox{\dimexpr\textwidth-\fboxsep-\fboxrule\relax}{\copyrighttext}}};
\end{tikzpicture}%
}

\def\BibTeX{{\rm B\kern-.05em{\sc i\kern-.025em b}\kern-.08em
    T\kern-.1667em\lower.7ex\hbox{E}\kern-.125emX}}
\begin{document}

\title{A Multi-Tenant Framework for \\ 
Cloud Container Services
}

\author{
\IEEEauthorblockN{Chao Zheng}
\IEEEauthorblockA{{\it Alibaba Group}\\ 
{\it c.zheng@alibaba-inc.com}}
\and
\IEEEauthorblockN{Qinghui Zhuang}
\IEEEauthorblockA{{\it Alibaba Group}\\ 
{\it zhijin.zqh@alibaba-inc.com}}
\and
\IEEEauthorblockN{Fei Guo}
\IEEEauthorblockA{{\it Alibaba Group}\\ 
{\it f.guo@alibaba-inc.com}}
}

\newcommand{\vc}{VC\xspace}
\newcommand{\eg}{e.g.,\ }
\newcommand{\ie}{i.e.\ }
\newcommand{\etal}{et al.~}
\newcommand{\todo}[1]{\textbf{\color{red}[todo: #1]}}

\graphicspath{{imgs/}}

\maketitle

\begin{abstract}
Container technologies have been evolving rapidly in the cloud-native era. 
Kubernetes, as a production-grade container orchestration platform, has been proven to be 
successful at managing containerized applications in on-premises datacenters. 
However, Kubernetes lacks sufficient multi-tenant supports by design,
meaning in cloud environments, dedicated clusters are required to
serve multiple users, i.e., tenants. 
This limitation significantly diminishes the benefits of cloud computing, 
and makes it difficult to build multi-tenant software as a service (SaaS) products using Kubernetes.
In this paper, we propose VirtualCluster, a new multi-tenant framework that
extends Kubernetes with adequate multi-tenant supports. Basically, VirtualCluster
provides both control plane and data plane isolations while sharing the underlying
compute resources among tenants. The new framework preserves the API
compatibility by avoiding modifying the Kubernetes core components. Hence, it can be easily
integrated with existing Kubernetes use cases.
Our experimental results show that the overheads introduced by 
VirtualCluster, in terms of latency and throughput, is moderate. 

\copyrightnotice

\end{abstract}


\section{Introduction}
\label{sec:introduction}

Modern application platforms empowered by container technologies
greatly simplify the processes of application development and management.
As a pervasive container orchestration platform, Kubernetes~\cite{k8s} provides
well-adapted resource abstractions, rich resource management capabilities,
and great extensibility for end users.
Almost all major cloud vendors provide container services~\cite{eks, aks, gke, ack} built on
top of Kubernetes. 
Kubernetes has been proven capable of support on-premises
clusters~\cite{k8s-case}, where users are trustworthy.
However, there is a debate about whether 
Kubernetes is suitable for cloud multi-tenant use cases, i.e., one 
cluster serves multiple customers, also known as tenants, or not.

Multi-tenancy is one of the essential attributes
of cloud computing for resource efficiency.
In the typical multi-tenant use case, the compute resources are usually abstracted, so tenants
would not need to know the details of the shared physical infrastructure.
Besides, tenants should not be aware of others' existence by all means
since they are not trustworthy.
We would argue that Kubernetes is barely satisfactory in supporting multi-tenancy.
First, to use Kubernetes properly, tenants have to understand the details of the 
node resources, such as the topology, the capacity, and the utilization.
More importantly, Kubernetes lacks strong isolation mechanisms.
In Kubernetes, an object is either cluster scoped or namespace scoped.
The namespace scoped Pod object is the most popular one, which encapsulates
the containers and describes their resource specifications.
Currently, to support multiple tenants in Kubernetes, a cluster administrator
has to use namespaces to group all tenant resources
and apply role-based access control (RBAC)~\cite{rbac} to limit the tenant accesses.
Kubernetes also supports network policy and Pod security policy to protect the
tenant containers. The above techniques are necessary but still
insufficient to satisfy the multi-tenant requirements in productions.

\begin{figure}[tbp]
\center
\includegraphics[width=0.6\columnwidth]{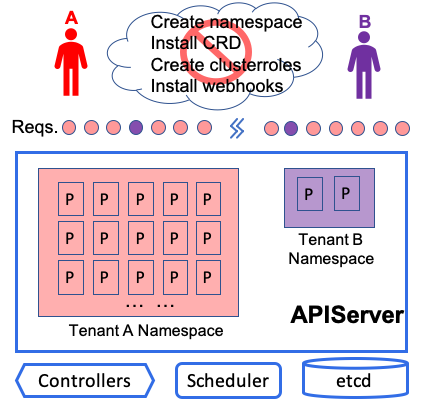}
\caption{The impact of sharing one Kubernetes control plane among multiple tenants.}
\label{fig:intro}
\end{figure}
The Kubernetes control plane is composed of an apiserver and a set
of built-in controllers. 
As illustrated in Figure~\ref{fig:intro}, 
allowing multiple tenants to share one apiserver will at least lead to the following problems:
\begin{itemize}
\item {\bf Performance interference.} When tenants simultaneously send requests to the apiserver,
performance abnormalities such as priority inversion, starvation, etc., may occur. In
the worst case, a buggy or overwhelming tenant (e.g., tenant A in Figure~\ref{fig:intro})
can completely crowd out others by issuing many queries against a large
number of resources. For instance, tenants may frequently query
all Pods in their namespace, making the requests from other tenants significantly delayed.
\item {\bf Lack of API supports.} Although using RBAC can prevent a tenant from accessing
the objects of others, some information cannot be protected without
proper API supports. For example, the namespace object is a cluster scoped object, while
the namespace \texttt{List} API cannot filter the result based on the tenant identity.
Once tenants are granted permission to list the namespaces to find their own ones, 
they can see all namespaces in the cluster, which might be problematic since the namespace name
may contain sensitive information.
\item {\bf Management inconvenience.} To avoid affecting others, tenants
usually have strict restrictions in operating cluster scope resources. As a result, tenants
cannot freely create namespaces, clusterroles, or install custom resource definitions (CRDs\cite{crd}),
and webhooks.
To work around such restrictions, tenants must go through rigorous negotiations
with the cluster administrator.
\end{itemize}

Kubernetes data plane does not fully support multi-tenancy either. Its
service discovery mechanism assumes a flat network model, meaning all Pods and
the node daemons are accessible from each other.
This assumption is completely broken in cloud multi-tenant environments where the tenant
containers typically connect to a virtual private cloud (VPC) to achieve network isolation.
As a result, native Kubernetes service APIs, such
as cluster IP type service, usually do not work in the public cloud without
vendor supports.

Fully supporting multi-tenancy in Kubernetes is challenging since
many core components need to be re-architectured. There are ongoing efforts to address some of the problems
in the community. For example, there is a proposal to implement priority and fairness for apiserver
requests ~\cite{KEP-fairness}. The multi-tenancy working group proposes a CRD to
allow namespace self-service creation~\cite{Multitenancy-WG}.
In enterprise products built on top of Kubernetes, the multi-tenancy support is usually 
implemented by encapsulating the Kubernetes apiserver with another layer and exposing 
a set of new APIs~\cite{openshift, anthos, wcp}.
Last but not least, one might create dedicated clusters for different
tenants, but the resulting low resource utilization would 
always be a concern.

In this paper, we propose VirtualCluster, a new framework that
addresses the above isolation problems in Kubernetes with specific design goals. 
The goals are:
\begin{itemize}
\item Supporting multi-tenancy in Kubernetes with full API compatibility, which
is the key to minimize the cost of integrating VirtualCluster with existing Kubernetes use cases.
\item Leveraging Kubernetes extensibility, i.e., CRDs, and avoiding modifying
the core components.
\item Sharing the node resources among tenants to maximize the resource utilization.
\end{itemize}
To achieve the above, in VirtualCluster, each tenant is assigned a dedicated 
Kubernetes control plane, which is referred to as a {\bf tenant control plane}. 
Tenants now can create cluster scope resources such as namespaces
or CRDs in their own control plane without affecting others, and most of the performance problems
due to sharing one control plane do not exist. The cluster that manages
the physical nodes now becomes the container resource provider, which is referred
to as a {\bf super cluster}. We developed a syncer controller to
populate objects from the tenant control planes to the super cluster, and update the object statuses back to the tenant control planes.
The syncer also ensures data consistency under the conditions of failures or races.
Kata sandbox container~\cite{kata} is used to provide a VM standard container runtime
isolation. The Kata agent running in each guest OS is slightly modified to work with an enhanced
kubeproxy to support Kubernetes cluster IP type of service.
We developed a virtual node agent to proxy all \texttt{log} and \texttt{exec} requests from the tenant control planes
to the Pods running in the super cluster.
With all the above, from a tenant's perspective, each tenant apiserver behaves 
like an intact Kubernetes with elastic cluster capacity.
The core VirtualCluster components are open-sourced in~\cite{vc}.

We have conducted experiments to evaluate the performance impact of the VirtualCluster
framework. The results show that in large-scale stress tests, the 
operation latencies of using VirtualCluster are comparable to baseline cases.
VirtualCluster also maintains a sustainable throughput.
For example, when creating ten thousand Pods in one hundred tenant control planes simultaneously, it took $\sim$23 seconds
to create all Pods. The same took $\sim$18 seconds when 
all Pods are created in the super cluster directly.
We have verified that VirtualCluster can pass all Kubernetes conformance
tests except one\footnote{The failed test requires the super cluster to use the subdomain name specified in
the tenant control plane. This cannot be supported in the current design.}.

In terms of the target use cases, VirtualCluster is particularly suitable for building a cloud 
container service or Kubernetes-based software as a service (SaaS) product.
Using VirtualCluster, a user can offload the burden of maintaining physical nodes and
only pay for the resources used by the workloads. Most of the existing Kubernetes plugins and operators can
be ported to VirtualCluster with almost zero integration efforts. Moreover, 
better compute resource utilization can be achieved when supporting multiple tenants.


\section{Background and Related Work}
\label{sec:backgroundandrelates}

Multi-tenancy has been a long-standing research topic in cloud computing. Previously,
researchers investigated how to improve isolation and security
in cloud environments~\cite{zhang2011cloudvisor, mosayyebzadeh2018secure,
shue2012performance, mosayyebzadeh2019supporting, chen2019enclavecache, zhai2016cqstr, kulkarni2018splinter}
for the control plane, networking, and storage. Some 
detection mechanisms were proposed to identify malicious tenants~\cite{rajasekaran2016scalable, hong2018go}.
Some studies focused on the resource management aspects in multi-tenant 
environments~\cite{dukic2019happiness, jeon2019analysis}
for better resource utilization. Many problems mentioned in previous work
do exist in Kubernetes, but the proposed solutions cannot be easily applied.
Some multi-tenant requirements need
architecture changes such as supporting nested namespaces,
supporting namespace scoped CRD, etc.
Without formal API supports, most of the existing Kubernetes multi-tenant solutions 
suffer from either API incompatibility or component version incompatibility
(due to modifying the component).

For example, virtual kubelet~\cite{vk} supplies Pods using
container services from cloud providers, such as AWS Fargate, instead of using worker nodes. 
Users can benefit from the new container techniques invented by the
cloud vendors, such as the sandbox runtime~\cite{aws-fargate}. 
However, virtual kubelet defines a simple
provider interface ($\sim$7 APIs vs. $\sim$25 CRI APIs implemented by
the kubelet) hence cannot fully support the Kubernetes Pod APIs, which inevitably
leads to usability issues in production.
There are hybrid cloud solutions, such as Google Anthos~\cite{anthos} and VMware
Project Pacific~\cite{wcp}, aim to provide serverless user experiences.
They address some of the multi-tenant problems by
introducing a new control plane or modifying the apiserver to leverage the 
underlying VM infrastructure. 
K3V from Rancher~\cite{k3v} is the closest open-source project compared to VirtualCluster.
In K3V, each tenant is assigned a modified Kubernetes distribution named K3S.
A controller is installed in each K3S, which copies tenant objects to a
cluster for Pod provision. 
K3V still relies on a modified apiserver and does not 
address any data plane isolation problems.
Loft~\cite{loft} is a commercial developer platform that has a feature
using a similar design in supporting multi-tenancy, i.e., each tenant is assigned a dedicated 
Kubernetes control plane. However, the tenant objects from multiple tenant
namespaces are copied to one namespace in the underlying cluster, which breaks the
Kubernetes API compatibility.
To the best of our knowledge, VirtualCluster is the first open-source effort
to support both control plane and data plane isolations natively in Kubernetes
without sacrificing the API compatibility. 

\begin{figure}[tbp]
\center
\includegraphics[width=0.75\columnwidth]{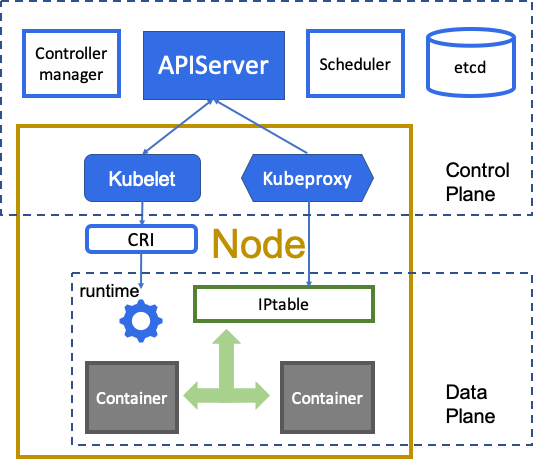}
\caption{Kubernetes architecture overview.}
\label{fig:k8s}
\end{figure}

\noindent {\bf Kubernetes Basics.} 
Kubernetes is an open-source container orchestration platform.
Figure~\ref{fig:k8s} presents a high-level overview of its architecture.
The Kubernetes control plane consists of several centralized components,
including an apiserver, a controller manager, and a scheduler. The user interacts with the
apiserver by operating various objects, and the Kubernetes
controllers ensure all objects reach their desired states eventually.
All object states are persisted in etcd~\cite{etcd} storage.
Kubernetes has dozens of built-in object types such as Pod, Service, Node etc., and it
allows users to create custom resource definition (CRD) to extend its capabilities.
It manages multiple worker nodes and uses a centralized scheduler to determine
where a Pod should run.
A few daemons
are installed in every worker node in which kubelet is the most important one.
Kubelet is responsible for container lifecycle management based on the
Pod specification stored in the apiserver. It interacts with node container runtime using
the container runtime interface (CRI). Container network/storage can be configured using network/storage
plugins. Kubernetes service discovery mechanism assumes that all container
network traffics goes through the host network stack. A node daemon, kubeproxy
~\cite{kubeproxy}, manipulates the host IPtable and configures the L3 forwarding rules
to route the requests to the service virtual IP to the service endpoints. Meanwhile, 
a service controller running on the control plane maintains the service virtual IP and its endpoints.

\begin{figure}[tbp]
\center
\includegraphics[width=0.9\columnwidth]{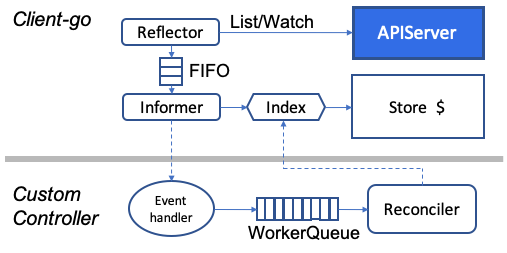}
\caption{The interactions between client-go components and custom controller.
Note that this figure is a simplified version of the figure in~\cite{sample}.}
\label{fig:controller}
\end{figure}

Kubernetes controllers manage the states of the cluster. Figure~\ref{fig:controller}
illustrates the primary workflow of a running controller utilizing the client-go library~\cite{client-go}.
A client-go reflector watches for the changes of a specific resource type 
in the apiserver. The changes are updated to a thread-safe store, i.e., a 
read-only cache, and sent to an informer's controller event handler. The
event handler sends the changed objects to a FIFO-style 
worker queue. The reconciler has multiple worker threads, each draining the worker queue and
performing the reconciling logic to reach the object's desired state. 
A worker thread can access
the read-only cache to retrieve an object's state, but all object updates are sent to the apiserver 
directly. In VirtualCluster, the resource syncer is a typical Kubernetes controller whose
goal is to keep the states of the synchronized objects consistent. 
More details are presented in Section~\ref{sec:objectsync}.

\section{Design}
\label{sec:virtualcluster}

In this section, we will present our design assumptions in Section~\ref{sec:assumption}, and 
the overall architecture
in Section~\ref{sec:system-arch}. We will describe 
the resource syncer design in detail in Section~\ref{sec:objectsync}, and
discuss the pros and cons of this design in Section~\ref{sec:benefits}.

\subsection{Assumptions}
\label{sec:assumption}
The proposed framework is designed for Kubernetes-based cloud container services.
We assume the following threat models and requirements:
\begin{itemize}
\item Tenant users are untrustworthy. They may generate harmful usage
patterns intentionally or unintentionally. A tenant cannot share objects with others in the
apiserver.
\item Containers are required to use tenant's virtual private cloud (VPC) 
through a vendor-specific network interface such as AWS elastic network interface~\cite{eni}, to achieve 
network isolation.
\item Containers are not safe. To prevent the containers from obtaining the node root privileges,
the service provider needs to run them using sandbox runtime.
\end{itemize}

\subsection{Architecture}
\label{sec:system-arch}

\begin{figure}[t]
\includegraphics[width=0.95\columnwidth]{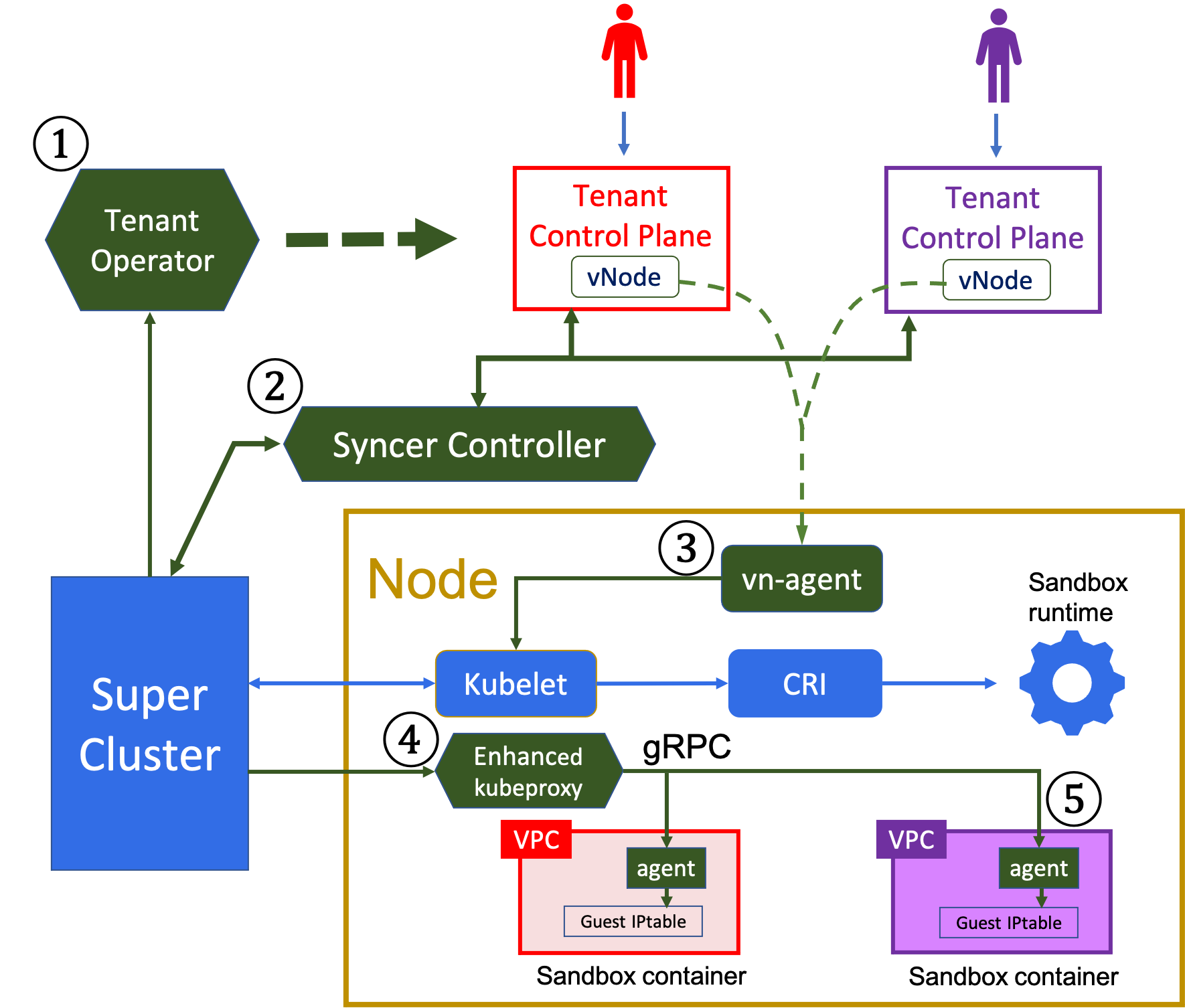}
\caption{VirtualCluster architecture overview. We omit the storage components in this figure since they are not affected
	in our design. For example, a dedicated etcd can be assigned to each tenant control plane.}
\label{fig:vc-arch}
\end{figure}

In Figure~\ref{fig:vc-arch}, the dark green components are newly introduced in VirtualCluster.
The super cluster is an upstream Kubernetes that manages the physical nodes
but behaves like a Pod resource provider. Tenant control planes own the sources of the truth
for the tenant objects and are responsible for maintaining their states.
We use Kata sandbox container runtime in each physical node to achieve VM standard 
runtime isolation. We will explain each component in detail.

\noindent \textcircled{\raisebox{-0.9pt}{1}} {\bf Tenant operator.} A VirtualCluster CRD,
referred to as VC, is defined to describe the tenant control plane 
specifications such as the apiserver version, resource configurations, etc.
VC objects are managed by the super cluster administrator. 
The tenant operator reconciles the states of VC objects whenever they are changed
to deal with the lifecycle events of the tenant control planes. 
 Note that a tenant control plane does not
need a scheduler since the Pod scheduling is done in the super cluster.
VC currently supports local mode and cloud mode to
provision tenant control planes.
In the cloud mode, the tenant operator leverages
the public cloud services such as AliCloud ACK~\cite{ack} or 
AWS EKS~\cite{eks}, to manage the control plane components.
It also stores the $kubeconfig$, the cluster access credential, of each
tenant control plane in the super cluster so that the syncer controller
can access all tenant control planes from the super cluster. Tenants are disallowed to access the 
super cluster.

\noindent \textcircled{\raisebox{-0.9pt}{2}} {\bf Syncer controller.} It only populates the
tenant objects used in Pod provision,
such as namespaces, Pods, services, secrets, etc., to
the super cluster, excluding all other control or extension 
objects. Note that all read requests to the synchronized objects
are served by the tenant apiservers, alleviating the
super cluster's pressure compared to the case where 
all tenants access the super cluster directly.
In Kubernetes, any namespace
scoped object's full name, i.e., \texttt{namespace/objectname} has to be unique. 
The syncer adds a prefix for
each synchronized tenant namespace to avoid name conflicts. The prefix is the concatenation of the owner VC's
object name and a short hash of the object's UID. 
The syncer does more than merely copying objects.
More details are discussed in section~\ref{sec:objectsync}.

\noindent \textcircled{\raisebox{-0.9pt}{3}} {\bf Virtual node agent.} In Kubernetes, kubelet can only 
register itself to one apiserver, i.e., the super cluster in VirtualCluster.
Hence, commonly used kubelet APIs such
as \texttt{log} and \texttt{exec} do not work for tenants since
the tenant apiserver cannot directly access the kubelet. We implement a 
virtual node agent (vn-agent) to resolve this problem, which runs in every
node to proxy tenants' kubelet API requests. More specifically, once a Pod is
scheduled in the super cluster, the syncer 
will create a virtual node object in the tenant apiserver.
To intercept the kubelet API requests, the virtual node points to
the vn-agent in the physical node instead of the kubelet.
When proxying the requests, vn-agent needs to identify the 
tenant from the HTTPS request because the tenant Pod has a different namespace
in the super cluster. The tenant who sends the request
can be found by comparing the hash of its TLS certificate with the one saved in
each VC object. The namespace prefix used in the super cluster
can be figured out after that.

\noindent \textcircled{\raisebox{-0.9pt}{4}} {\bf Enhanced kubeproxy.}
In Kubernetes, the cluster IP type service defines a routing policy to access 
a set of endpoints (i.e., Pods) inside the cluster.
The routing policy is enforced by a kubeproxy daemon, which updates the host IPtable whenever
the service endpoints change.
This mechanism is broken when containers are connected to a virtual private cloud (VPC)
because the network traffics might completely bypass the host network stack
through a vendor-specific network interface.
To enable cluster IP type service in such an environment, we enhance the kubeproxy
by allowing it to directly inject or update the network routing rules in 
each Kata container's guest OS. 
More specifically, the Kata agent (\textcircled{\raisebox{-0.9pt}{5}}) running inside the guest OS
opens a secure gRPC connection with the kubeproxy through which
the service routing rules can be applied in the IPtable of the guest OS.
The changes to the kubeproxy are moderate. It needs to 
watch for the Pod creation events and coordinate with a Pod initcontainer, which is 
run ahead of any workload container and checks for the IPtable update progress,
to ensure the routing rules are injected before the workload containers start.

\subsection{Resource Syncer}
\label{sec:objectsync}

\begin{figure}[t]
\includegraphics[width=0.95\columnwidth]{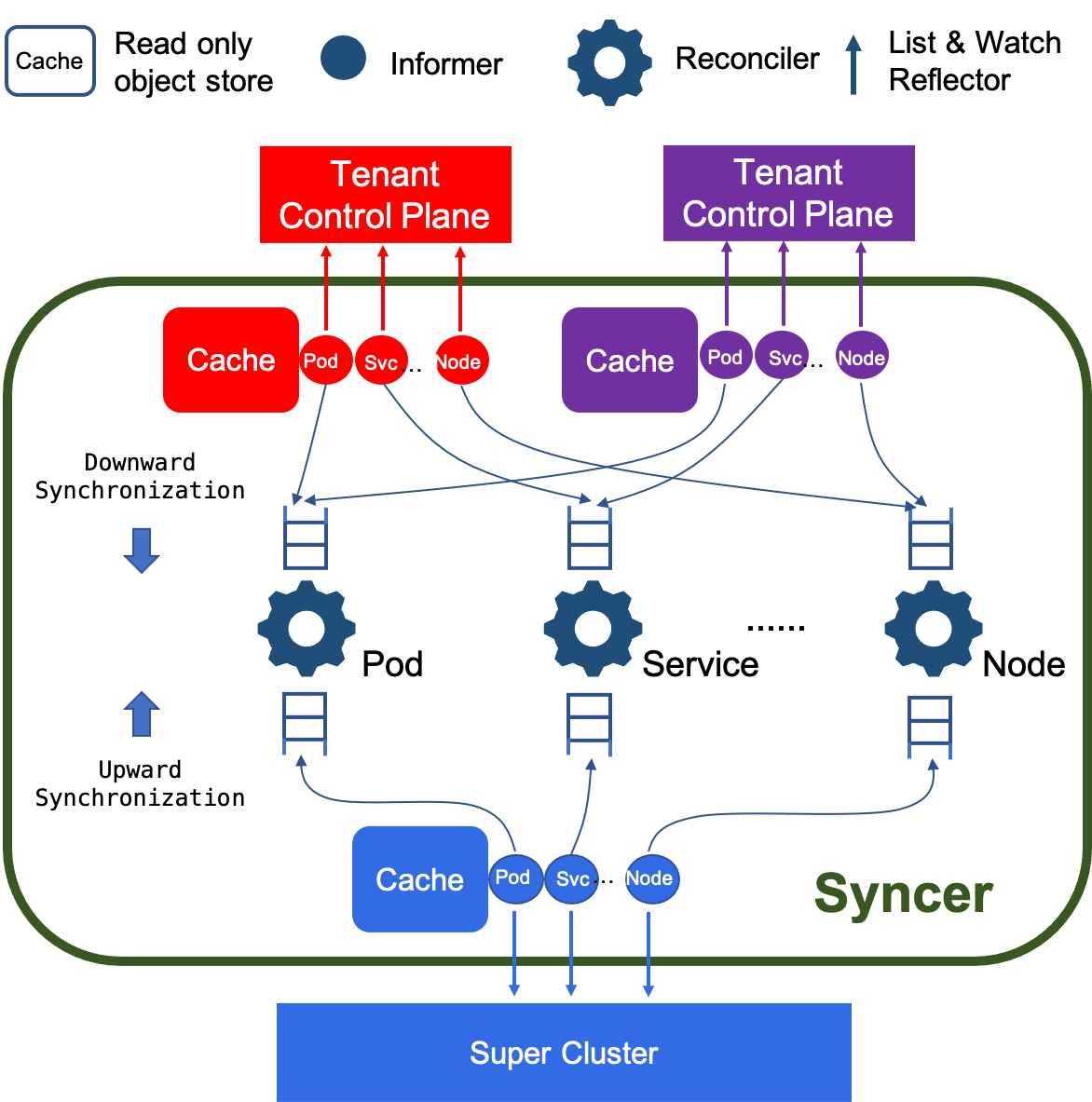}
\caption{Resources syncer components. The caches are read only object stores used in
	client-go informers. There are per resource reconcilers which perform
	downward and upward synchronizations.}
\label{fig:syncer}
\end{figure}

The syncer could be installed for every tenant like the design in \cite{k3v}.
We choose a centralized design in VirtualCluster due to a few reasons.
First, the object create, update or delete operations issued by the tenants are generally infrequent.
It would be a waste of resources had a syncer been installed per-tenant basis.
Secondly, the syncer needs
to fetch the watched objects' states from the super cluster to its informer cache whenever
itself or the super cluster apiserver restarts. If there are too many of them, when
the super cluster apiserver restarts, the object \texttt{list} requests from the syncers could quickly flood
the super cluster and make it unserviceable. Therefore, in VirtualCluster, 
one syncer instance serves many tenant
control planes. If the syncer
or the super cluster apiserver restarts, the super cluster objects' states are fetched only once.
In Section~\ref{sec:evaluation}, we will show that this centralized design will not cause
scalability issues for latencies.

\begin{figure}[t]
\centering
\includegraphics[width=0.9\columnwidth]{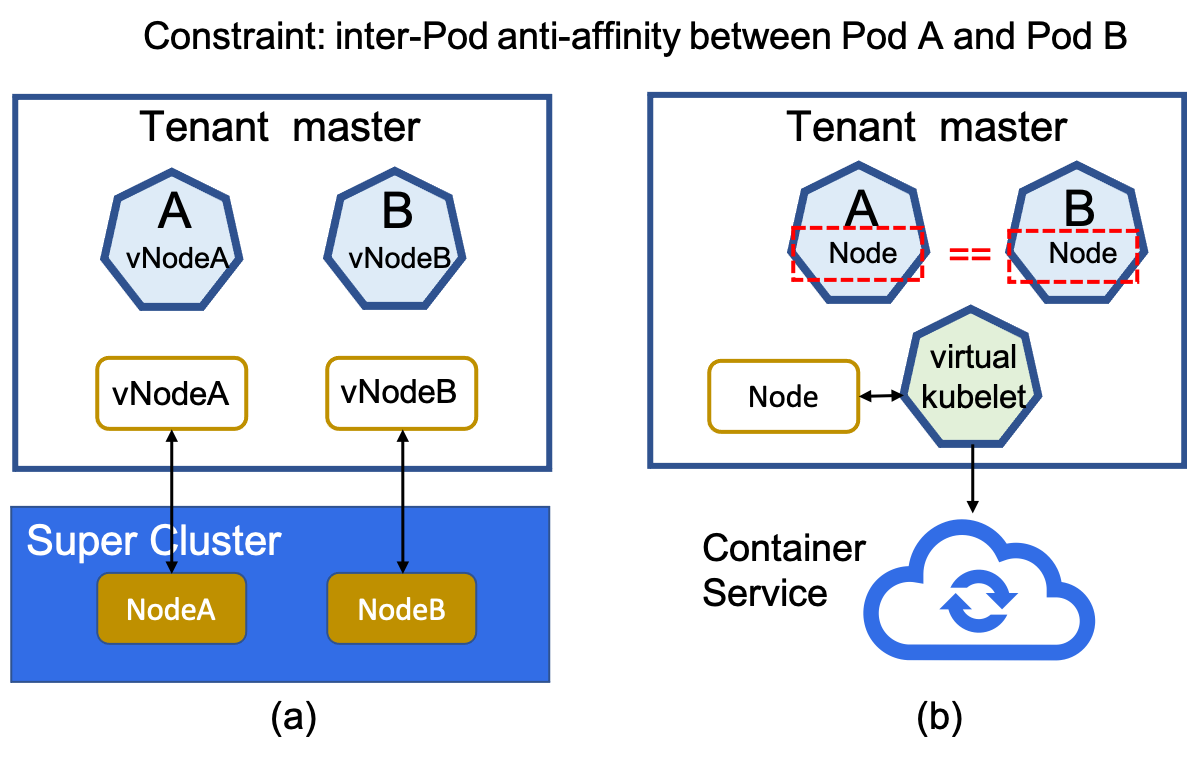}
\caption{Comparison between a vNode in VirtualCluster and a virtual kubelet.}
\label{fig:vcvsvk}
\end{figure}

The syncer currently synchronizes twelve types of resources between the super cluster
and the tenant control planes, which are sufficient to ensure compatible behaviors from a tenant's perspective. 
Figure~\ref{fig:syncer} presents the internal structure of the syncer controller.
Based on the sources of the truth of the
synchronized objects, the syncer either populates the states (e.g., Pod specifications) from tenant
control planes to the super cluster, i.e., {\bf downward} synchronizing, or back populates the states (e.g., Pod statuses) 
from the super cluster to a tenant control plane, i.e., {\bf upward} synchronizing, using the per resource reconcilers. 
The state comparisons are made against the super cluster and tenant control plane
informer caches to avoid intensive direct apiserver queries, assuming the client-go reflectors work reliably.
From Figure~\ref{fig:syncer}, we can see that all tenant informers send
the changed objects to a shared downward FIFO worker queue, which can lead to a 
well-known queuing unfairness problem for tenants.
To eliminate the potential contention, we extend the standard client-go worker queue
with fair queuing support. Specifically, we add per tenant sub-queues and use the
weighted round-robin scheduling algorithm~\cite{wrr} to dispatch tenant objects to the downward worker queue.
As a result, none of the tenants would suffer from significant object synchronization delays, preventing
starvation.

Kubernetes controllers follow an eventual consistency model. Hence the object states in the informer cache
can be inconsistent with the states in the apiserver in a short period (usually a few milliseconds).
This temporal inconsistency may occasionally introduce races in the syncer. For example, 
an object might have been deleted in the apiserver while the syncer
is handling the object's update event. Although the syncer is resilient to such races, 
some states could be permanently inconsistent under rare failure conditions.
To deal with this issue, the syncer will periodically scan the synchronized objects
and remediate any state mismatch by resending the object to the 
worker queue again. This design significantly reduces the complexity of recovering
inconsistencies caused by various rare reasons. 
In Section~\ref{sec:evaluation}, we will show that the cost of periodic scan
is moderate.

The syncer controller manages all virtual node objects in the tenant control planes.
The physical node heartbeats will be broadcasted to all virtual nodes periodically.
The binding associations between the tenant Pods and the virtual nodes are tracked in the syncer as well.
Once a virtual node has no binding Pods, it will be removed from the tenant control plane 
by the syncer. Each virtual node object represents a real physical node
in the super cluster from a tenant's perspective. 
The one-to-one mapping between a virtual node and a real node 
is a unique abstraction that preserves all Kubernetes node semantics.
The user experience in VirtualCluster is quite different compared to that of
using a virtual kubelet, which typically connects to a cloud container service,
not a real node. For example, as illustrated in Figure~\ref{fig:vcvsvk}, assuming there is a constraint that
Pod A and Pod B cannot run in the same host, i.e., an inter-Pod anti-affinity rule,
this constraint is correctly represented in VirtualCluster since two Pods are bound to
different vNodes (Figure~\ref{fig:vcvsvk}(a)). However, in the case of using virtual kubelet
(Figure~\ref{fig:vcvsvk}(b)),
two Pods are bound to the same virtual kubelet node object, which will cause user confusion since
the user has no idea whether the constraint has been enforced or not.

We assign multiple worker threads for downward and upward worker queues to speed up the synchronizations.
The syncer's memory footprint is determined
by the total size of the synchronized objects in the informer caches, which is
dominated by the size of the Pod objects in most cases. There
are dynamic memory allocations in the worker queues as well. Note that 
each tenant control plane has Kubernetes built-in rate limit control enabled,
and more importantly, the client-go worker queue has the capability of deduplicating the
incoming quests, the memory consumptions of the
worker queues are unlikely to grow infinitely.

\subsection{Discussions}
\label{sec:benefits}
The benefits of this design are straightforward. By providing dedicated
control planes to tenants, the classic noisy neighbor problems  
due to sharing a single control plane are largely mitigated. The blast radius of security
vulnerability is also limited. If a tenant triggers a control plane 
security issue, only that tenant is the victim. A tenant has full permissions to operate 
the tenant control plane, hence gains the same user experiences as if using a dedicated
upstream Kubernetes.
However, better isolations can be costly. The required resources for 
running all tenant control planes grow as the number of tenants increases. There is an advantage such 
that Kubernetes follows a thin server, thick client design pattern. 
All extended capabilities are implemented in client controllers. Hence the
resource requirements of running core Kubernetes components are usually small and stable.
VirtualCluster has limitations. For example, there exist other shared components in the 
super cluster uncovered in this paper, such as the local image store, local volumes, etc. 
Their isolations are out of the design scope of VirtualCluster. In addition, 
other aspects of multi-tenancy besides isolation such as security, availability
depend on the upstream Kubernetes capabilities and are not addressed in VirtualCluster. 
Besides, VirtualCluster cannot support cases where tenants need to install plugins in the shared worker nodes.

\section{Evaluation}
\label{sec:evaluation}

\noindent{\bf Environment.} We set up the super cluster using a 1.18 Kubernetes cluster that consists of two worker
nodes. Each worker node is a bare-metal 4-sockets Intel Xeon 8163 machine with 96 cores and 328GB memory.
As we need to measure the resource usage of the syncer controller, to avoid potential interferences from
other components, we deployed the syncer controller in one worker node, and 
all tenant control planes in another worker node. Each tenant control plane used a dedicated etcd.
All control planes are connected to the same VPC using a high-speed virtual switch.
Due to the limited resources, we installed one hundred virtual kubelets in the super cluster to simulate a cluster with one hundred
nodes running a large number of Pods. Note that 
the Pod creation's latencies reported in our experiments exclude the time spent on pulling images
and constructing containers in real nodes because of using virtual kubelet.
These are static overheads and not affected by VirtualCluster at all.


\begin{figure*}[t]
{\centering
    \begin{tabular}{c|c|c}
{\centering {\bf 25 Tenants}} & {\centering {\bf 50 Tenants}} & {\centering {\bf 100 Tenants}} \\
\includegraphics[width=0.31\textwidth]{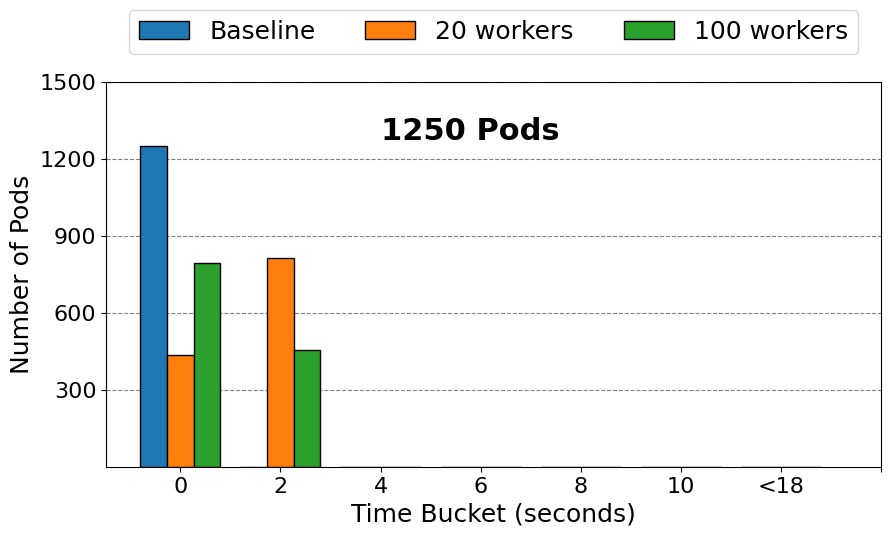} & \includegraphics[width=0.31\textwidth]{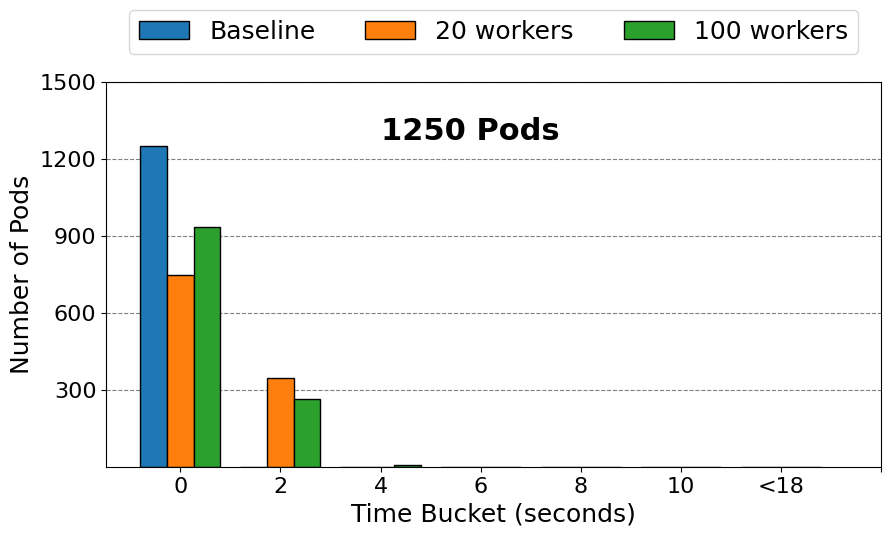} & \includegraphics[width=0.31\textwidth]{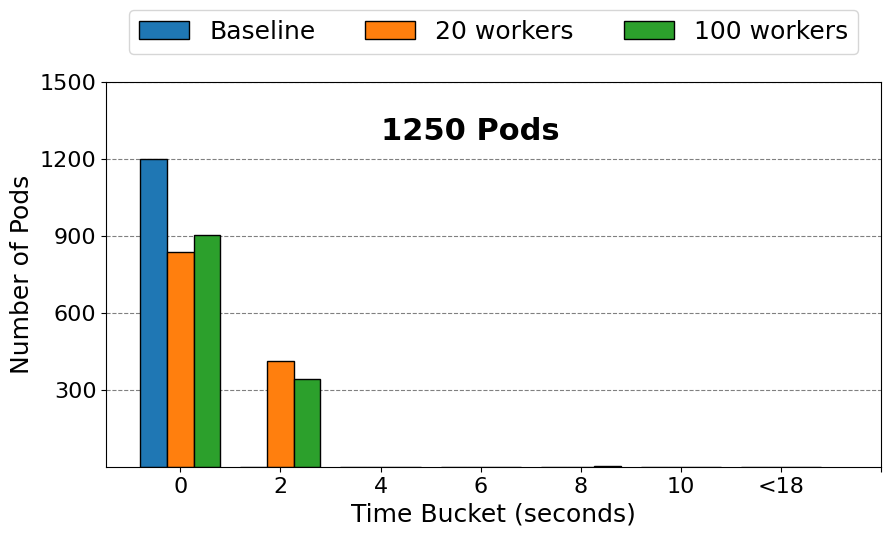} \\
\includegraphics[width=0.31\textwidth]{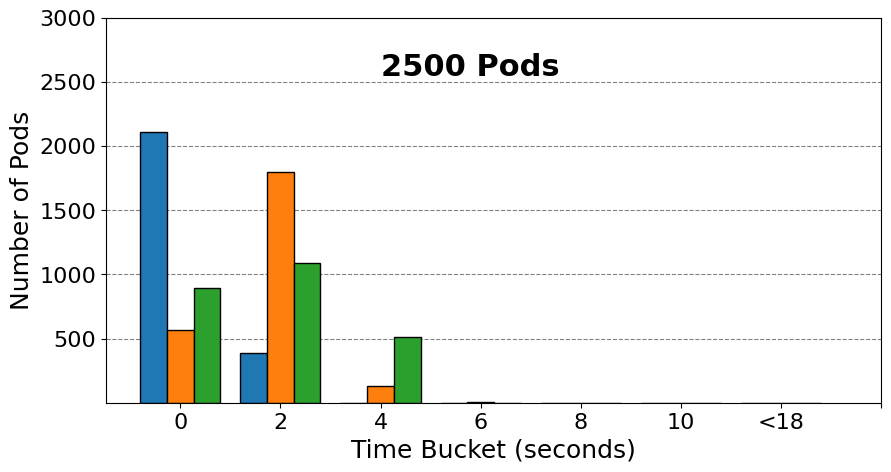} & \includegraphics[width=0.31\textwidth]{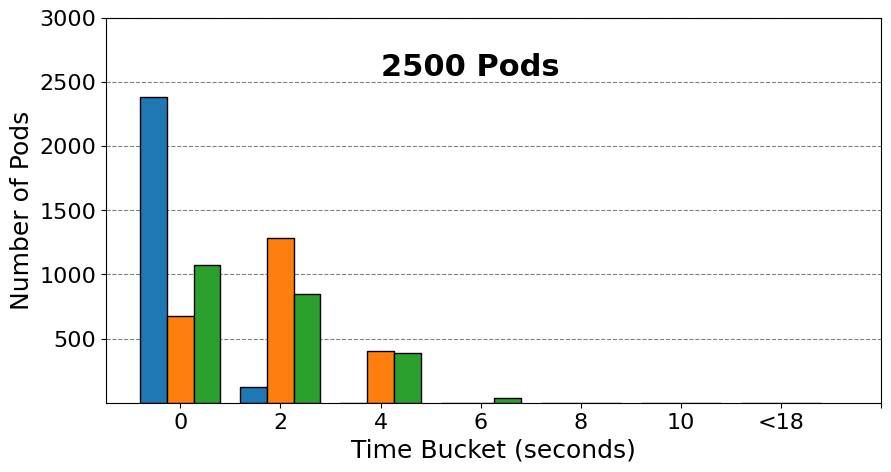} & \includegraphics[width=0.31\textwidth]{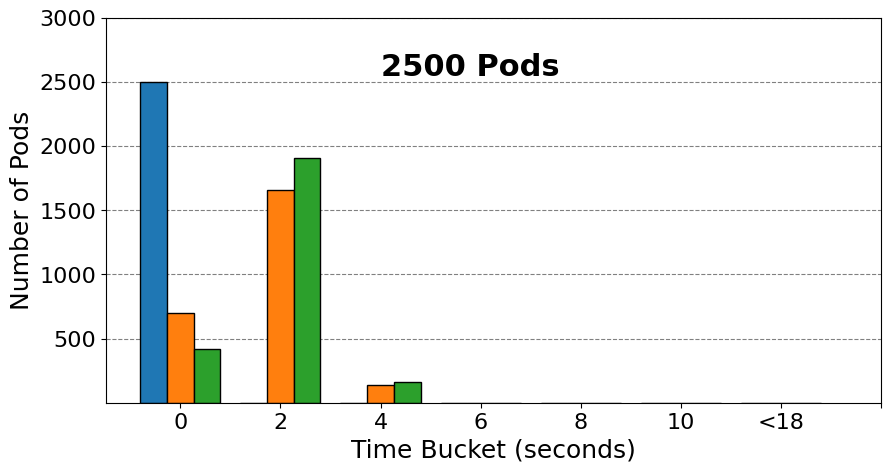} \\
\includegraphics[width=0.31\textwidth]{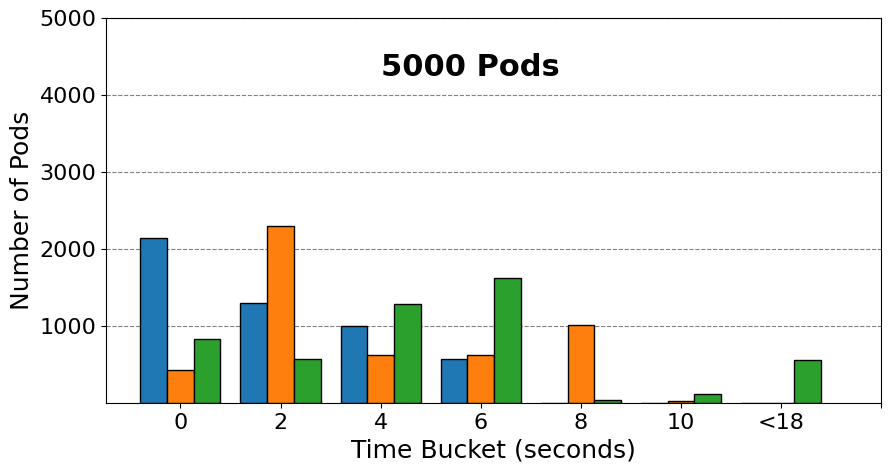} & \includegraphics[width=0.31\textwidth]{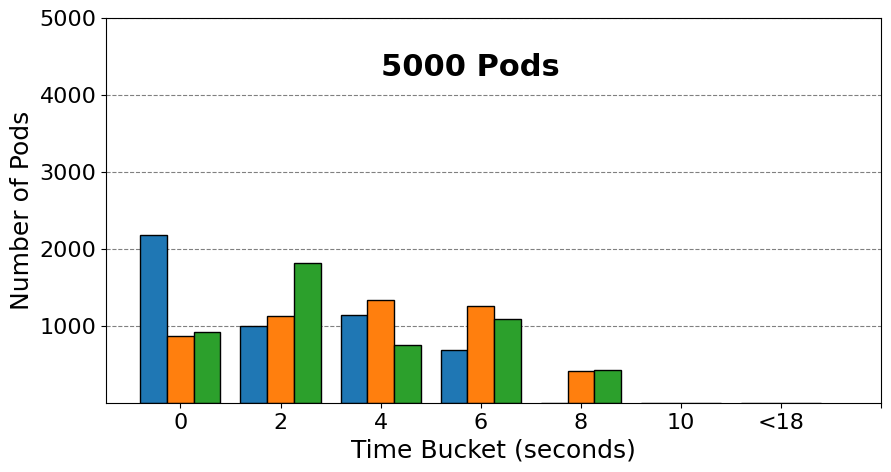} & \includegraphics[width=0.31\textwidth]{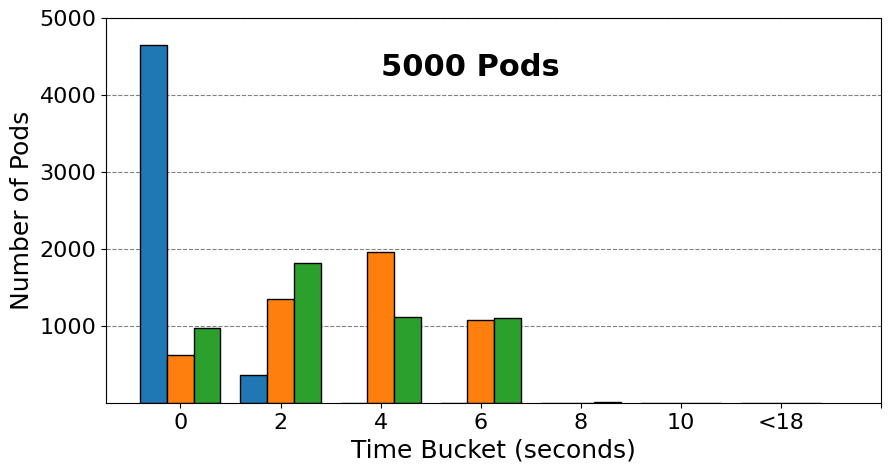} \\
\includegraphics[width=0.31\textwidth]{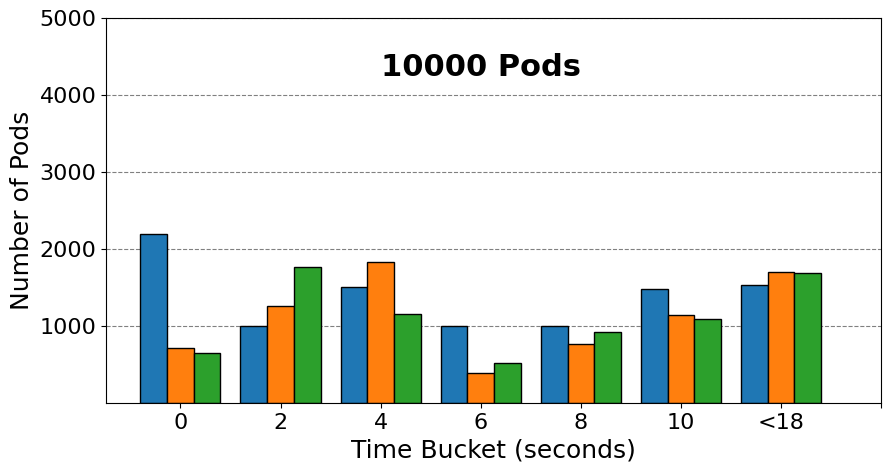} & \includegraphics[width=0.31\textwidth]{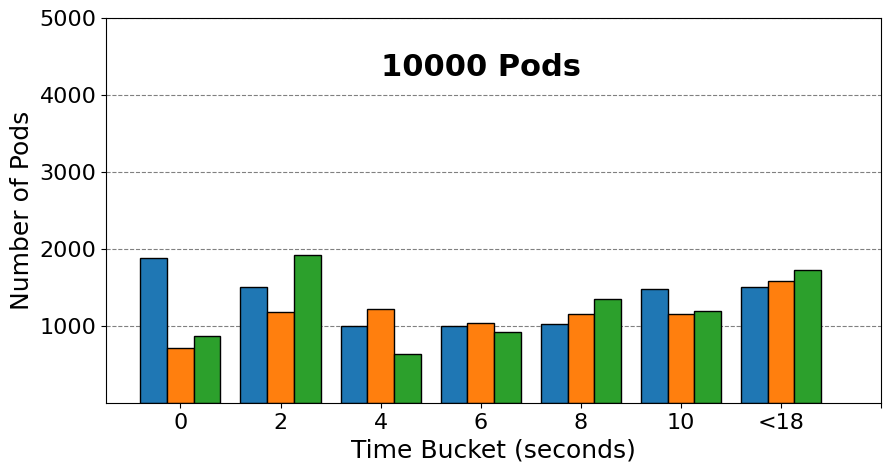} & \includegraphics[width=0.31\textwidth]{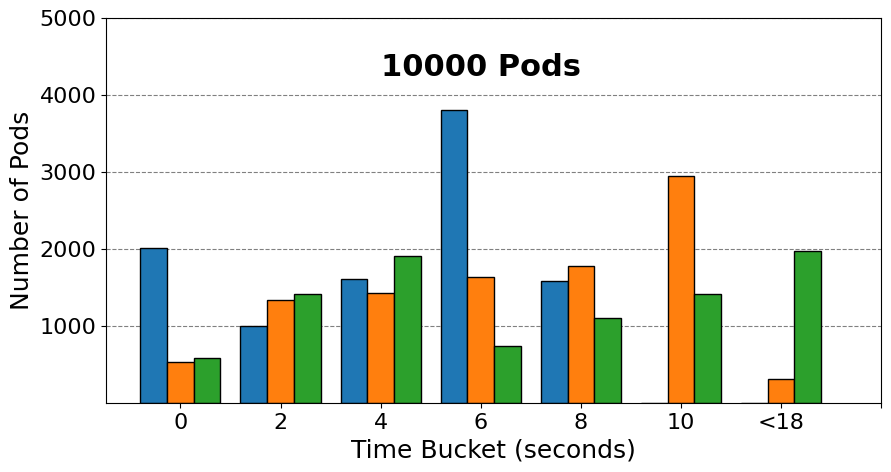} \\
\end{tabular}

\captionsetup{justification=centering}
\caption{The histograms of Pods creation time with different number of tenants, different number of Pods and different number of downward worker threads in the syncer.}
\label{fig:pods-creation-latency}}
\end{figure*}



\noindent{\bf Workload.} 
Since VirtualCluster uses a dedicated control plane for each tenant, the isolation between tenants is solid. 
Therefore, our evaluation focused on the performance impact of the framework. 
We chose the end-to-end Pod creation time as the major performance metric due to the following reasons:

\begin{itemize}
\item The queuing delay introduced by the syncer is the primary performance concern when VirtualCluster is
under heavy loads. 
\item The Pod object, as one of the primary objects in Kubernetes, has arguably the most complicated schema and 
many other objects were invented to serve it.
The performance of operating Pod objects will be highly representative.
\item Pod creation triggers a complicated workflow involving
other major Kubernetes components such as scheduler and kubelet. It is in the
critical path of application deployment. Hence its performance is often highlighted.
\end{itemize}
We believe VirtualCluster will perform equally or even better when handling 
other objects such as services, namespaces, secrets, etc.

When VirtualCluster is under normal loads, e.g., tens of requests per second, we found the syncer 
added one or two milliseconds delays, which are negligible in typical Kubernetes use cases.
We developed a load generator
that created a large number of Pods simultaneously in all tenant control planes to stress the system.
In the super cluster, each virtual kubelet runs a \texttt{mock} Pod provider, which marks
all Pods scheduled to the virtual kubelet ready and running instantaneously.
The Pod creation time was measured as the difference between the tenant Pod creation timestamp
and the timestamp that the Pod's condition is updated as ready in the tenant, including all queuing delays
and the object synchronization overheads added by the syncer.
We evaluated a few baseline cases in which the load generator
sent all requests to the super cluster directly to make comparisons. The Pod
creation time in the baseline was calculated based on the Pod creation timestamp and
the Pod's ready condition timestamp.

\subsection{The impact of the syncer controller on latency}

First, we considered three factors that could impact the syncer performance in terms of
latency. They are the number of created Pods, the number of tenants, and
the number of downward worker threads.
We conducted twelve cases by varying the number of each factor, and the results
are presented in Figure~\ref{fig:pods-creation-latency}. In each case, the number of Pods 
created in each tenant is equal.
In contrast, for the baseline, the load generator used the same number of threads as 
the number of tenants to submit Pods to the super cluster directly. For each configuration,
we repeated the tests several times and calculated the average for evaluation.

\begin{figure}
\centering
\includegraphics[width=0.8\columnwidth]{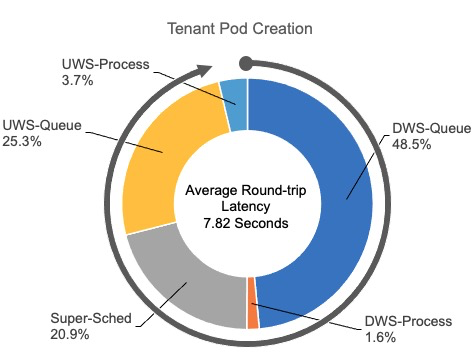}
\caption{The breakdown of the average Pod creation round-trip latency when creating ten thousand Pods in one hundred tenant control planes simultaneously.}
\label{fig:latency-breakdown}
\end{figure}

We compared the Pod creation time histograms against the histograms
of the baseline cases, and the results are presented in Figure~\ref{fig:pods-creation-latency}. 
A concentrated histogram means stable performance, and a flat histogram indicates high-performance variations.
Figure~\ref{fig:pods-creation-latency} shows that using VirtualCluster
does not significantly lengthen the Pod creation time.
The majority of the operations had latencies within
the baseline latency range. For example, when using one hundred tenants and
twenty worker threads, the 99\% percentile latencies were 3 (vs. 1 in the baseline) seconds,
4 (vs. 2 in the baseline) seconds, 8 (vs. 8 in the baseline) seconds, 14 (vs. 8 in the baseline) seconds
when creating 1250, 2500, 5000, and 10000 Pods, respectively.
It is interesting to see that the baseline cases expressed noticeable performance
variations with a high number of created Pods.
We found that the scalability bottleneck of the super cluster was the scheduler.
The default Kubernetes scheduler has a single queue, and it schedules Pod sequentially. Therefore, we have seen the scheduler throughput
peaked at a few hundred Pods per second in our experiments.
The queuing delay in the scheduler can slow down the Pod creation process under a high Pod churn rate,
which also explained another observation such that increasing
the number of downward worker threads did not help reduce the latencies at all.
As we will illustrate later, the time spent on the syncer's downward reconciling loop is trivial (Figure~\ref{fig:latency-breakdown}).
Using twenty worker threads was enough to push the super cluster to reach the upper limit of the scheduling throughput. However,
the number of upward worker threads did affect the latency (not shown in the figure)
since the tenant control plane had no bottleneck in handling object status updates.
Therefore, we set a high default
number of one hundred upward worker threads and a low default number of twenty
downward worker threads in the syncer.
Those default numbers were used in the rest of the experiments unless otherwise mentioned.

In Figure~\ref{fig:pods-creation-latency}, we can also observe that the number of
tenants did not impact the latency for the same amount of
created Pods. Intuitively, the syncer performance should be affected as the number
of tenants increased since the syncer needed to create more sub-queues
to support fair queuing and the complexity of the weighted round-robin algorithm, used in sub-queue dequeue, is $O(n)$, where n is the number of sub-queues.
However, in our experiments, all tenants were assigned the same weight. Hence the algorithm
effectively became a standard round-robin algorithm with $O(1)$ complexity.
If the weights of tenants were different\footnote{Currently, VirtualCluster
does not support custom weight for different tenants, which is part of our future work.},
we would expect the latency not to be significantly affected unless the number of tenants
was large.

To understand where the time was spent during Pod creation in VirtualCluster,
we divided the Pod creation latency into five phases in chronological order:
1) The time spent in the downward worker queue (DWS-Queue);
2) The downward synchronization time (DWS-Process);
3) The time spent in the super cluster until the Pod is marked as ready and running (Super-Sched);
4) The time spent in the upward worker queue (UWS-Queue);
5) The upward synchronization time (UWS-Process).
Figure~\ref{fig:latency-breakdown} presents the average latency breakdown in the case
of creating ten thousand Pods in one hundred tenants.
In the figure, we can see that the delays in the two syncer worker queues contribute $\sim$75\% of the latency on 
average, 48.5\% by the downward worker queue and 25.3\% by the upward worker queue, respectively.
The time spent in the downward and upward synchronizations is negligible.
The scheduling delay (21\%) in the super cluster is remarkable due to the reasons explained above.
\begin{table}
\centering	
\begin{tabular}{|l|c|c|c|c|c|}
\hline
	\backslashbox{Phase}{Bucket} & {[}0, 2{]} & {[}2, 4{]} & {[}4, 6{]} & {[}6, 8{]} & {[}8, 10{]} \\ \hline
DWS-Queue       & 2935      & 2663      & 1626      & 1998      & 778        \\ \hline
DWS-Process  & 10000     & 0         & 0         & 0         & 0          \\ \hline
Super-Sched & 3607      & 6393      & 0         & 0         & 0          \\ \hline
UWS-Queue      & 2798      & 6870      & 332       & 0         & 0          \\ \hline
UWS-Process  & 10000     & 0         & 0         & 0         & 0          \\ \hline
\end{tabular}
\caption{The time bucket counts of each Pod creation phase in the same case of Figure~\ref{fig:latency-breakdown}.
The bucket unit is second.}
\label{tb:latency-histo}
\end{table}
Table~\ref{tb:latency-histo} presents the detailed bucket counts of each Pod
creation phase considering all Pods. As we can see, the delay variations in all
phases are small besides the DWS-Queue phase, in which the burst requests start to accumulate.

Overall, the experiment results show that a centralized syncer can handle burst Pod creation requests with
small to moderate added delays. Note that such delays may not be noticeable
in practice when considering the time spent in the real node during
Pod creation (e.g., the time to pull container image).
In addition, since the super cluster scheduler can be the bottleneck of the syncer scalability,
adding more worker threads or more syncers cannot improve the latency effectively.
Lastly, the syncer is stateless and can be scaled out to support a huge number of tenants if needed.

\begin{figure}
\centering
 \begin{tabular}{cc}
	 \includegraphics[width=0.4\columnwidth]{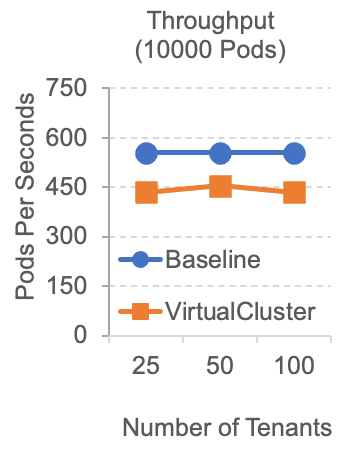} & \includegraphics[width=0.4\columnwidth]{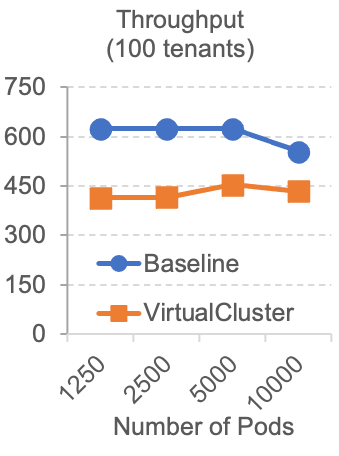}\\
	 (a) & (b) \\
 \end{tabular}
\caption{The Pod creation throughput comparisons 
by fixing the number of Pods (a) and fixing the number of tenants (b).}
\label{fig:throughput}
\end{figure}

\subsection{The impact of the syncer controller on throughput}
Next, we evaluated the VirtualCluster throughput by calculating the number of created Pods per second and
the results are presented in Figure~\ref{fig:throughput}.
From Figure~\ref{fig:throughput}(a), we can see that the number
of tenants does not affect the throughput for the same number of created Pods.
VirtualCluster introduced a constant $\sim$21\% throughput degradation.
The lower throughput is expected since a few critical sections in the syncer, 
such as the worker queue enqueue or dequeue cannot be parallelized.
The lock contentions in the syncer could downgrade the throughput.
Figure~\ref{fig:throughput}(b) shows that the throughput is roughly
constant for VirtualCluster, but becomes lower for the baseline cases as the number of
Pods increases. The maximal throughput degradation is $\sim$34\%.
Note that adding more syncers might improve the overall throughput
by reducing per syncer lock contentions.
However, it is not preferable due to a few reasons: 1) Using one syncer still
achieved sustainable throughput regardless of the number of tenants and
the number of Pods; 2) The operation latency is more important from a tenant's perspective
compared to the control plane throughput.

\begin{figure}
{\centering
\includegraphics[width=.9\columnwidth]{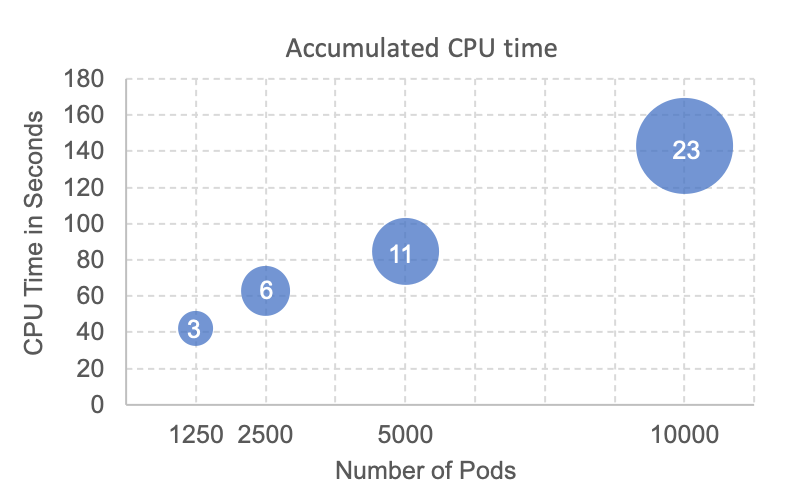} \\
\includegraphics[width=.9\columnwidth]{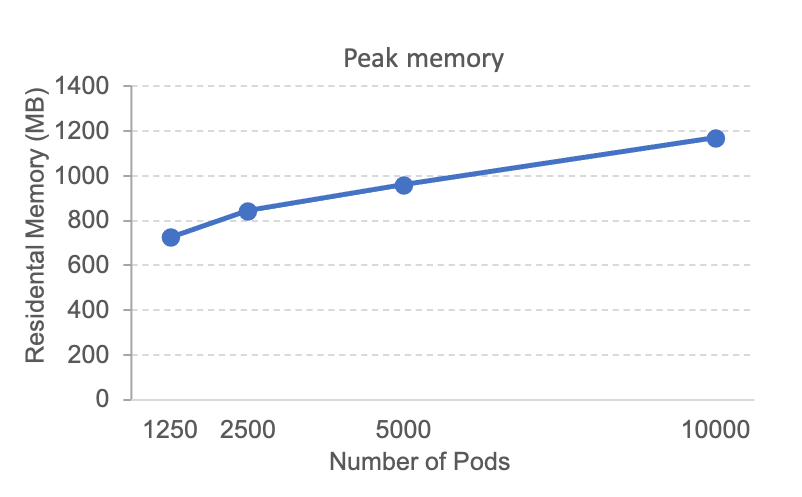} 
	\caption{The resource usages of the syncer controller. The CPU usage was measured as the accumulated process CPU time.
	The number in each circle represents the process wall-clock time (in seconds) to which the size of each circle is proportional. 
	The memory usage was measured as the process's peak resident set size (rss).}
\label{fig:syncer-usage}}
\end{figure}

\subsection{The overhead of the syncer controller}
The available compute resources for the syncer can affect its performance when it is busy.
In our experiments, we did not set a resource limit for the syncer. It would
be interesting to analyze the resource usages in those intensive test cases built for benchmarking
purposes.
Figure~\ref{fig:syncer-usage} presents the CPU and memory usages of the syncer controller
in our experiments.
As expected, the resource usages increase almost linearly as the number of Pods increases.
The average number of consumed CPUs can be estimated
by calculating the division of the accumulated CPU time by the process wall-clock time.
For example, in the ten thousand Pods case, the syncer roughly consumed $\frac{138}{23}$ i.e., six CPUs during
the experiment, which is far beyond the requirements for normal cases.
Usually, a CPU limit of one to two CPUs is recommended for the syncer.
The syncer's peak memory usage is around 1.2GB in the ten thousand Pods case.
The peak memory growth rate is roughly 40KB per Pod, which is estimated by calculating
the ratio of the curve.
The major memory consumers in the syncer are the informer caches.
One tenant object has at least two copies in the
syncer, one in the informer cache of the tenant control plane and another in the super cluster
informer cache. The syncer worker queues also consume memory when they grow, but the queued request's size is usually small (a few bytes), 
and the queues would not grow infinitely because of deduplication.
We have also examined the syncer restart performance and found that
it took less than twenty-one seconds to initialize all informer caches with one hundred tenant control planes
and ten thousand Pods, which is reasonably fast because syncer restart
would be rare. Besides, we have also measured the overhead of the periodic scanning threads
in the syncer. The number of parallel scanning threads was equal to the number of tenants, and the scan interval
was set to one minute. We found that it took less than two seconds to finish scanning 10000 Pods on average. 

\begin{figure}[t]
\centering
\captionsetup{justification=centering}
\captionsetup[subfigure]{justification=centering}

\begin{subfigure}[b]{\linewidth}
\centering
\includegraphics[width=\linewidth]{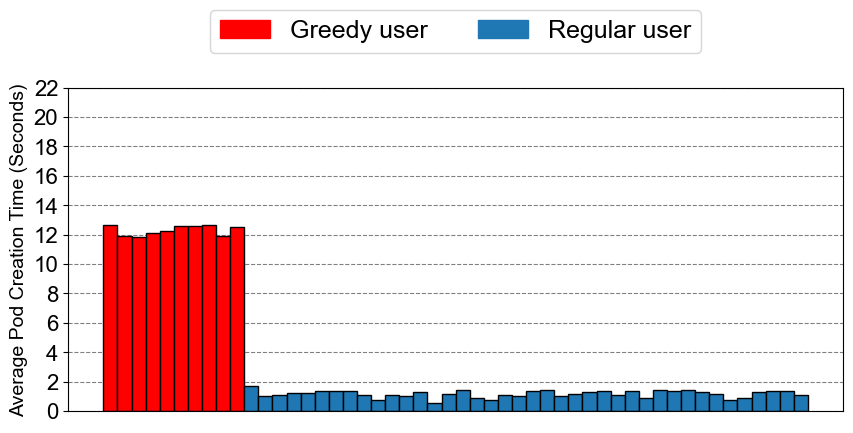}
\caption{Fair queuing enabled}
\label{fig:with-fair-queue}
\end{subfigure}

\begin{subfigure}[b]{\linewidth}
\centering
\includegraphics[width=\linewidth]{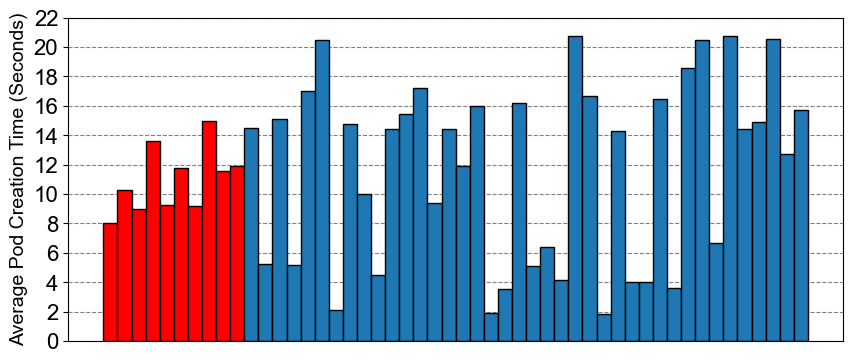}
\caption{Fair Queuing disabled}
\label{fig:without-fair-queue}
\end{subfigure}

\caption{The average Pod creation time of each tenant with and without fair queueing enabled.}
\label{fig:fair-queuing}
\end{figure}

\subsection{The impact of fair queuing on fairness}

In previous experiments, the load generator sent an even number of Pod creation requests to each
tenant, assuming they have the same usage pattern. To evaluate the impact
of the fair queuing mechanism in the syncer on performance, we divided the tenants into two groups:
ten greedy users and forty regular users. Each greedy user issued nine hundred Pod creation requests concurrently,
while each regular user sent ten Pod creation requests sequentially. 
All tenants had the same weight. Hence, the syncer should ensure that
the regular users' Pods would not be affected by the greedy users' burst requests from the aspect of fairness. As shown in
Figure~\ref{fig:fair-queuing}(a), under the help of the fair queueing mechanism, the average Pod creation time for all regular users
was small (less than two seconds), and all greedy users suffered
from much higher average Pod creation time. To make a comparison, we repeated the
experiment with fair queuing disabled in the syncer. As shown in Figure~\ref{fig:fair-queuing}(b),
the shared worker queue caused severe contentions, and the creations of many regular users' Pods were significantly
delayed due to the burst requests from the greedy users.
Note that without a centralized syncer, it would be challenging to implement fair queuing.
For example, if each tenant had a  syncer, we had to rely on the super cluster's apiserver to
provide quality of service (QoS) when serving concurrent requests. Unfortunately,
Kubernetes has no mature QoS support for user access controls yet~\footnote{Kubernetes does provide user-based
request rate limit controls, but it is hard to set proper limits for tenants in practice.}.

\subsection{The impact of the enhanced kubeproxy on latency}

Lastly, we evaluated the enhanced kubeproxy performance using a different methodology.
In the experiment, we created Pods in one real worker node in the super cluster instead of
using a virtual kubelet, i.e., creating thirty Pods in one worker node using
kata container runtime and connected to a VPC.
We also created one hundred artificial services beforehand so that the enhanced kubeproxy
would inject one hundred routing rules into each guest OS before the workload containers started.
We found that the extra latency caused by injecting those rules was $\sim$1 second on average, including the gRPC cost and the time to update the
IPtable. The time to scan all thirty Pods rules was around three hundred
milliseconds, which lengthened the periodic
reconciling loop's execution time in the kubeproxy.
Overall, the cost of supporting the cluster IP type of service in VirtualCluster
is small.

\section{Future work}
\label{sec:future}

Several cloud SaaS products have adopted the VirtualCluster framework.
We have realized that VirtualCluster can be further improved in at least the following aspects
in order to accommodate more use cases:
\begin{itemize}
\item {\bf Synchronizing CRDs}. The syncer controller currently 
only synchronizes the built-in Kubernetes resources used for Pod provision.
However, the super cluster may offer extended scheduling capabilities by
introducing new CRDs. For example, there exist quite a few scheduler
plugins for running artificial intelligence (AI) or big data workloads
in Kubernetes using new CRDs. A tenant user cannot use the
extended scheduling capability unless the syncer starts to synchronize the required
CRD from the tenant control plane. Therefore, adding CRD support in the syncer 
is a legitimate request and in our roadmap.

\item {\bf Reducing the cost of running tenant control planes}.
For a few tenants, the resources used for the tenant control planes 
would not be a concern. However, the cost could be a blocking factor if the
number of tenants reaches thousands or more. How to reduce the
tenant control plane resources, especially for idle tenants, is challenging.
Since Kubernetes is moving towards supporting memory swapping for the running
Pod~\cite{mem-swap}, one possible solution is to allow memory overcommitment
in the nodes that run the tenant control planes 
and swap the idle tenant control plane memory out.
However, the above idea requires a sophisticated design to
make proper tradeoffs between the performance and the cost. 

\item {\bf Supporting multiple super clusters}.
Since VirtualCluster hides the underlying super cluster capacity to the tenants,
the ability of autoscaling is the key for the super cluster to provide reliable
services to the tenants. In cases where worker nodes cannot be
automatically added to or removed from a super cluster, supporting
multiple super clusters is an option to break through the 
capacity limitation of a single super cluster. 
Note that this request is different from the use cases that
Kubernetes federation~\cite{kubefed} targets, where the users 
explicitly know the states of all managed clusters. In VirtualCluster,
the users would not be aware of multiple super clusters,
making the solution more robust and retaining a consistent user experience.
\end{itemize}

Last but not least, some VirutalCluster components such as the
vn-agent could be simplified by leveraging newly proposed Kubernetes
features. We will continue to work closely with the upstream community.

\section{Conclusion}
\label{sec:conclusion}

In this paper, we propose VirtualCluster, a new multi-tenant framework that
provides complete control plane isolations among tenants and allows them
to share the underlying compute resources. It enhances Kubernetes data plane
to support cluster IP type of service in VPC environments.
VirtualCluster complements Kubernetes by working around its design limitations
in multi-tenancy. From a tenant's perspective, VirtualCluster presents an intact 
Kubernetes cluster view by preserving full API compatibility. The experimental 
results show that VirtualCluster introduces small to moderate overheads
in terms of operation latency and throughput.
Overall, VirtualCluster framework can be easily integrated with 
most of the existing solutions, and we believe it has great potential to support
more multi-tenant use cases in cloud computing.

\bibliographystyle{IEEEtran}
\bibliography{vc-icdcs}
\end{document}